\documentclass[aps,amsmath,amssymb,floatfix,twocolumn]{revtex4}

\usepackage{bm}
\usepackage{graphicx}
\usepackage{amsbsy}
\usepackage{dcolumn}
\usepackage{color}

\newcommand{\be}{\begin{equation}}
\newcommand{\ee}{\end{equation}}
\newcommand{\bea}{\begin{eqnarray}}
\newcommand{\eea}{\end{eqnarray}}
\newcommand{\eqnm}[1]{Eq.\ (\ref{#1})}
\newcommand{\fignm}[1]{Fig.\ \ref{#1}}

\begin{document}

\title{Microscopic optical potential for exotic isotopes from chiral effective field theory}
\author{J.\ W.\ Holt$^{1,2}$, N.\ Kaiser$^3$ and G.\ A.\ Miller$^1$}
\affiliation{$^1$Physics Department, University of Washington, Seattle,
  WA}
\affiliation{$^2$Cyclotron Institute and Dept.\ of Physics and Astronomy, Texas A\&M University, College Station, TX}  
\affiliation{$^3$Physik Department, Technische Universit\"{a}t M\"{u}nchen, Garching, Germany}

\begin{abstract}
We compute the isospin-asymmetry dependence of microscopic optical model potentials from 
realistic chiral two- and three-body interactions over a range of resolution scales $\Lambda \simeq 
400-500$\,MeV. We show that at moderate projectile energies,
$E_{\rm inv} = 110 - 200$\,MeV, the real isovector part of the optical potential changes sign,
a phenomenon referred to as isospin inversion. We also extract the strength and energy 
dependence of the imaginary
isovector optical potential and find no evidence for an analogous phenomenon over the 
range of energies, $E \leq 200$\,MeV, considered in the present work. Finally, we compute for the 
first time the leading corrections to the Lane parametrization for the isospin-asymmetry dependence of 
the optical potential and observe an enhanced importance at low scattering energies.
\end{abstract}

\maketitle

{\it Introduction -- }
The structure and dynamics of neutron-rich
nuclei are key inputs for modeling neutron stars, core-collapse supernovae and 
$r$-process nucleosynthesis 
\cite{Steiner:2004fi,surman09,brett12,goriely11,sekiguchi11,bauswein13,pinedo12,roberts12}. 
Elucidating the properties of highly isospin-asymmetric nuclear matter is therefore a priority
in low-energy nuclear science research and a major motivation for the development of 
next-generation radioactive ion beam (RIB) facilities. Microscopic many-body methods
\cite{binder13,hagen14,hebeler15} with chiral two- and three-body forces 
have been successful in describing the bound-state properties of neutron-rich
matter. Complementary and consistent nuclear reaction models are under development 
\cite{holt13b,toyokawa15,navratil16}, and of these, global optical potentials aim to address 
the broadest theory needs for interpreting RIB scattering experiments and simulating 
$r$-process nucleosynthesis. In fact, current modeling of the strong $r$-process 
favors a cold scenario in binary neutron star mergers
\cite{pinedo12,roberts12,korobkin12,bauswein13,wanajo14,rrapaj15}, 
where mass transfer to highly neutron-rich isotopes occurs 
and freeze-out is achieved more rapidly, which enhances the importance of radiative neutron capture 
processes in determining the final abundance pattern of $r$-process elements \cite{mumpower12}.

Although global phenomenological optical potentials \cite{becchetti69,varner91,koning03} 
are well constrained close to the valley of nuclear stability, their predictive power for reactions 
involving exotic neutron-rich isotopes is not well understood.
Elastic scattering data has been used in the past to parametrize local optical potentials in 
specific regions of the nuclear chart off stability, however, the most exotic and low-intensity 
radioactive ion beams require thick-target experiments that provide quality inelastic data only
\cite{riley05}. This motivates the need for accurate microscopic optical model potentials
and investigations of their energy and isospin-asymmetry, $\delta_{np} = (N-Z)/A$, dependence. 
Identifying energy regimes in which the leading linear $\delta_{np}$ term is
dominant can then be valuable for extrapolating existing phenomenological 
potentials away from stability.

The aim of the present work is to employ high-precision chiral two- and three-nucleon forces to study
the real and imaginary volume components of the nucleon-nucleus optical model potential
far from the valley of stability. The dependence on the isospin asymmetry of 
the target nucleus is traditionally taken to be linear and isovector in character, a parametrization 
known as the Lane form \cite{lane62}. The isoscalar components of the optical potential are then 
independent of $\delta_{np}$. We revisit these assumptions and find that subleading
terms in $\delta_{np}$ (which are isoscalar and isovector for even and odd powers of 
$\delta_{np}$, respectively) can be important for highly neutron-rich nuclei and particularly
at the low energies most relevant for nuclear astrophysical phenomena. We study
the energy dependence of these terms up to $E \simeq 200$\,MeV in anticipation of 
future experimental investigations of exotic isotope reactions at RIB facilities.

A phenomenon of particular interest is isospin inversion, whereby the real isovector optical potential 
is expected to change sign from positive at low energies to negative at higher energies. In the vicinity 
of isospin inversion, subleading terms proportional to $\delta_{np}^2$ may 
become relevant and probe novel isospin physics. The interplay between 
intermediate-range attractive contributions to the nucleon-nucleon interaction and short-range repulsive 
contributions can give rise to a change in sign of the isoscalar optical potential at projectile 
energies greater than $E \sim 250$\,MeV, which has been observed in calculations 
with microscopic nucleon-nucleon potentials 
\cite{jeukenne76}. The isovector contribution to the optical potential, arising from $\pi$-meson exchange 
and $\rho$-meson exchange in traditional one-boson exchange models, has a stronger relative energy 
dependence. In the past, semi-microscopic and microscopic optical potentials have been 
constructed from mean field theory \cite{li04,chen07,shen09} and realistic nucleon-nucleon 
interactions \cite{jeukenne77a,jeukenne77b,bombaci91,zuo99,kaiser02,dalen05} 
respectively, and there is significant disagreement regarding the energy dependence of the isovector
components. To date there are therefore no strong constraints on the kinematic transition region associated with
isospin inversion.

In microscopic many-body theory the nucleon-nucleus optical model potential is identified with the 
on-shell nucleon self-energy $\Sigma(\vec r, \vec r^{\, \prime};E)$ \cite{bell59}.
In the present study we compute the nucleon self-energy at second order in many-body perturbation 
theory employing as a starting point high-precision 
nuclear interactions derived from chiral effective field theory \cite{epelbaum09,machleidt11}.
Chiral nuclear potentials with momentum-space cutoffs $\Lambda 
\lesssim 450$\,MeV \cite{coraggio07,coraggio13,coraggio14} exhibit very good perturbative 
behavior (comparable to renormalization-group evolved potentials
\cite{bogner03,bogner05,hebeler11}) and we also consider a potential employing a 500\,MeV 
cutoff \cite{machleidt11} that is used to give a conservative estimate of theoretical uncertainties
associated with nonperturbative dynamics and variations in the resolution scale.
Previous work has shown that chiral low-momentum potentials provide a good description of the symmetric 
nuclear matter saturation energy and density \cite{coraggio14}, the incompressibility and 
isospin-asymmetry energy \cite{holt12}, and the critical endpoint of the liquid-gas phase transition \cite{wellenhofer14,holt13}. The present approach to nuclear scattering is therefore consistent
with nontrivial constraints from nuclear structure.

{\it Isospin-asymmetry dependence of optical potentials -- }
In the optical model for nucleon-nucleus scattering, the complicated many-body problem 
associated with multiple scattering through two- and three-body forces is replaced by an 
(elastic-scattering) equivalent complex-valued single-particle potential:
\be
V(\vec r, \vec r^{\, \prime};E) = U(\vec r, \vec r^{\, \prime};E) + i W(\vec r, \vec r^{\ \prime};E),
\label{omp}
\ee
which in general is both non-local and energy-dependent. The imaginary part in 
\eqnm{omp} accounts for the presence of open inelastic scattering channels. 
Phenomenological optical potentials are often taken 
to be local and energy-dependent. The real and imaginary parts contain volume components
proportional to Woods-Saxon densities $f_j(r) = 1/(1+e^{(r-R_j)/a_j})$:
\begin{equation}
U(r;E) = -\overline{U}(E)f_r(r), \hspace{.1in}
W(r;E) = -\overline{W}(E)f_i(r),
\label{pomp}
\end{equation}
where the parameters $\overline U_0(E), \overline W_0(E), R_{r,i}$ and $a_{r,i}$ 
vary smoothly with the mass number $A$ of the target nucleus and the 
projectile energy $E$.

\begin{figure}
\includegraphics[width=8.5cm]{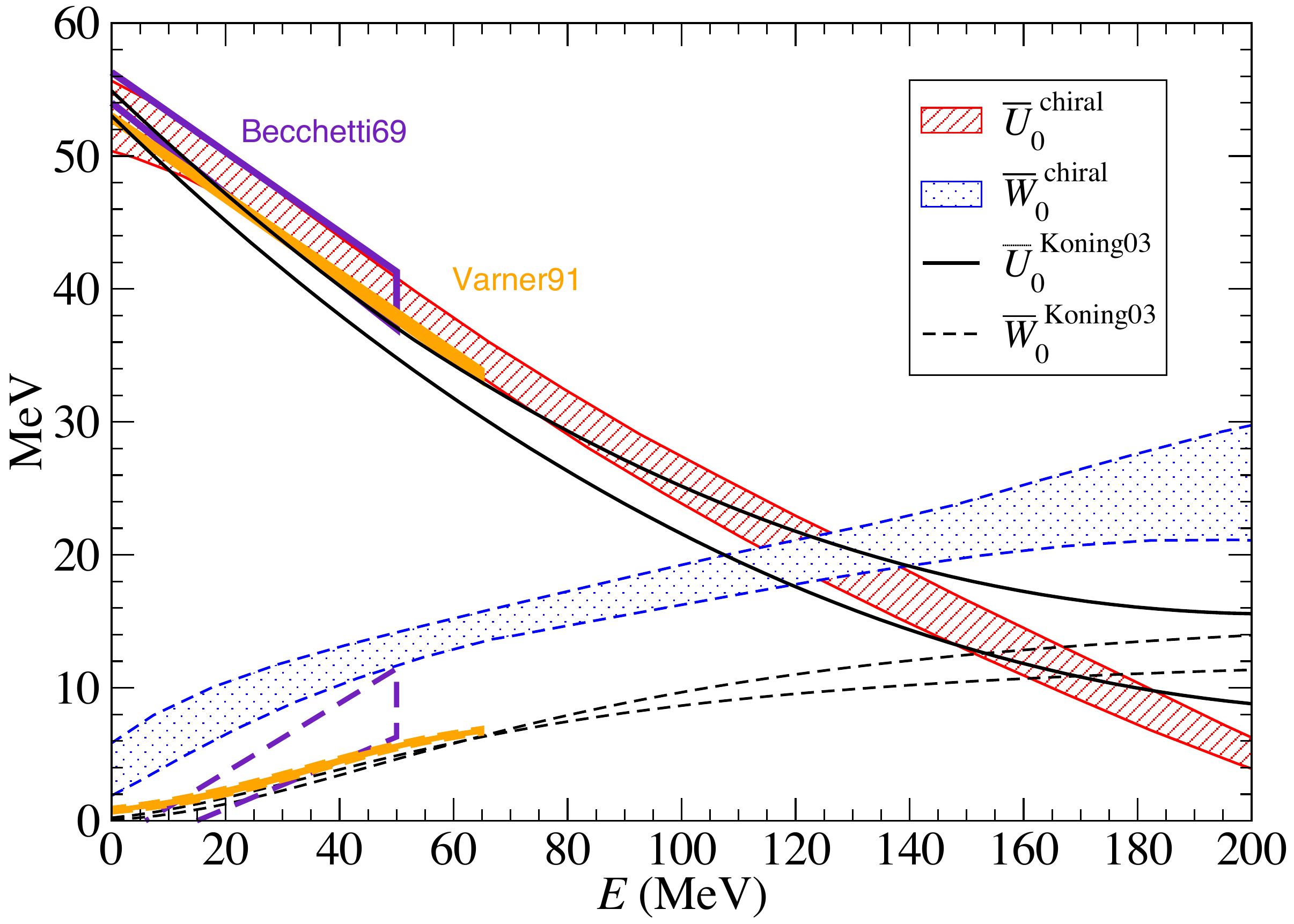}
\caption{(color online) Energy dependence of the real and imaginary parts of the microscopic 
optical model potential from chiral two- and three-body forces for symmetric nuclear matter
at saturation density $\rho_0$. Shown
for comparison are the global phenomenological potentials of Ref.\ \cite{becchetti69,varner91,koning03}.}
\label{OptPotScal}
\end{figure}

Recently chiral two- and three-nucleon forces have been used to compute the real and 
imaginary volume components of the optical potential for isospin-symmetric nuclear systems 
\cite{holt13b,toyokawa15}. Although the strength and energy dependence of the real component 
was found to be in good agreement with modern phenomenological parametrizations \cite{koning03}, 
the absorptive strength of the imaginary part from microscopic nuclear potentials was about a factor 
of two larger than that from phenomenology. We note that at low energies the phenomenological
imaginary part is surface peaked and vanishes in the infinite matter limit. Microscopic many-body 
theory in the local density approximation attributes the surface imaginary part 
to the nonlinear density dependence of the imaginary volume component.
In Fig.\ \ref{OptPotScal} we show the real and imaginary
parts of the optical potential at nuclear matter saturation density $\rho_0 = 0.16$\,fm$^{-3}$ from
chiral nuclear potentials compared to the global fit in Ref.\ \cite{koning03}. The phenomenological 
``Koning'' bands are obtained by varying the mass number over the range $A = 50-150$ and should not
be interpreted as an uncertainty. On the other hand, the error band associated with the microscopic 
calculation comes from varying the resolution scale over the range $\Lambda = 414-500$\,MeV. 
In Fig.\ \ref{OptPotScal} we have included for comparison also global optical potential parametrizations 
\cite{becchetti69,varner91} valid at lower energies with associated uncertainty estimates.
In contrast to the results reported in Ref.\ \cite{holt13b}, the single-particle energies entering in the
second-order perturbative calculation are computed self-consistently via 
$e(q) = q^2/2M + {\rm Re}\,\Sigma(q,e(q))$
rather than from the effective mass plus energy shift parametrization $e(q) \simeq q^2/2M^*+\Delta$, 
which smears out the enhancement of the momentum-dependent effective mass at the Fermi
surface \cite{bertsch68}. From Fig.\ \ref{OptPotScal} we observe that the microscopic real volume 
component remains nearly linearly dependent on the incident energy beyond $E=100$\,MeV,
in contrast to the Koning analysis in Ref.\ \cite{koning03}. Nevertheless, the two approaches 
are consistent within uncertainties over a wide range of energies.

For scattering on neutron-rich nuclei, the dependence of the optical potential on the isospin asymmetry 
$\delta_{np}$ is crucial. The standard Lane form
\be
U = U_0 + \frac{\vec \tau \cdot \vec T}{A}U_I,
\ee
where $\vec \tau$ and $\vec T$ are the isospin operators for the projectile and target nucleus 
respectively, is widely used in both phenomenological and microscopic calculations. 
The Lane parametrization relates the elastic proton-nucleus, neutron-nucleus, and quasi-elastic 
charge-exchange processes. For elastic scattering the Lane form reduces to
$U = U_0 - U_I \delta_{np} \tau_3,$
where $\tau_3 = \pm 1$ (for protons and neutrons, respectively) 
is the isospin quantum number of the incident nucleon.

\begin{center}
\begin{figure}
\begin{center}
\includegraphics[height=3.1cm]{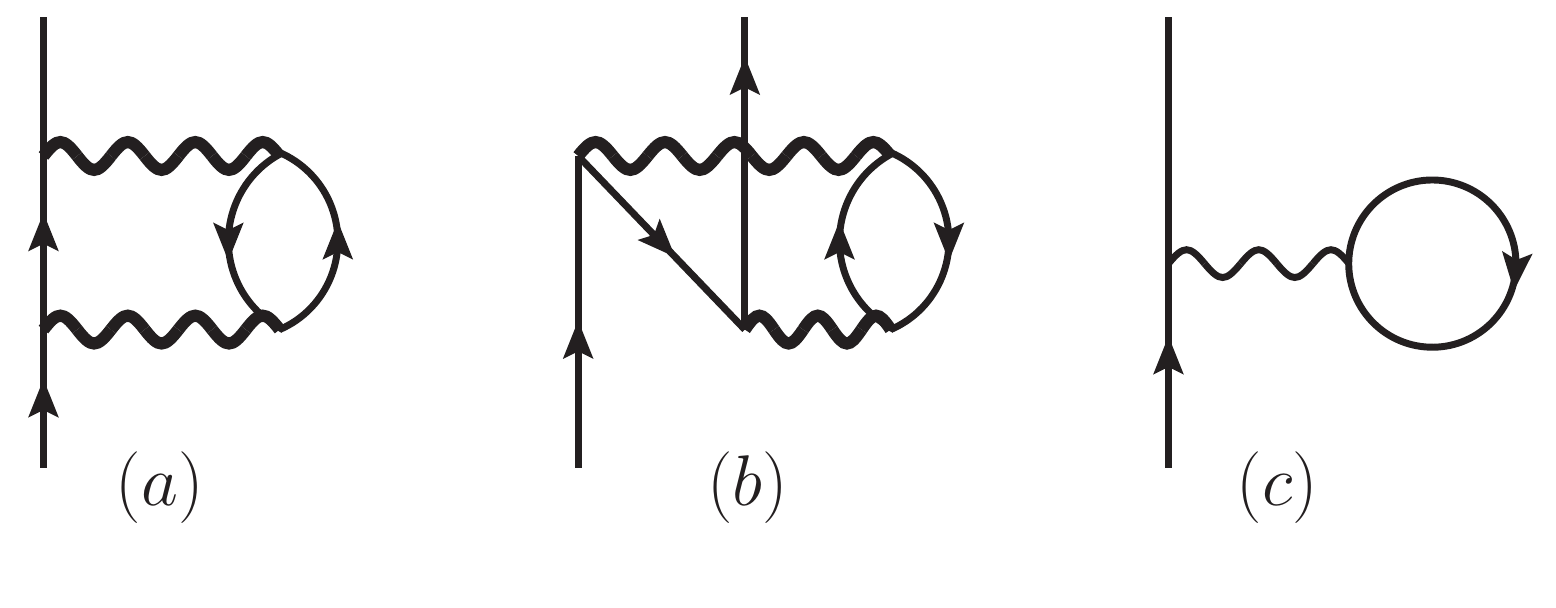}
\end{center}
\vspace{-.5cm}
\caption{Diagrams contributing to the proton and neutron self energies $\Sigma_{p,n}(q,\omega;k_f,\delta_{np})$ 
at first and second order in 
perturbation theory. The wavy line represents the 
antisymmetrized two-nucleon interaction $\bar V_{2N}$ and the thick wavy lines on the second-order diagrams 
represent the sum of the
free-space two-body force and the density-dependent NN interaction from Refs.\ \cite{holt09,holt10}.}
\label{op2nf}
\end{figure}
\end{center}

Here we consider a more general expansion of the isospin asymmetry dependence:
\begin{equation}
U = U_0-U_I\, \tau_3 \delta_{np}+ U_{I\!I} \delta_{np}^2+{\cal O}(\delta_{np}^3)\,.
\label{gexp}
\end{equation}
The Hartree-Fock contribution $\Sigma^{(1),2N}_{p,n}(q;k_f,\delta_{np})$ from 
two-body forces, shown diagrammatically in \fignm{op2nf}(c), is obtained by generalizing
the results of Ref.\ \cite{holt13b} and has the form
\be
\Sigma^{(1),2N}_t(q;k_f,\delta_{np}) = \sum_{1} \langle \vec q \, \vec h_1 s s_1 t t_1 | \bar V_{2N} | \vec q \,
\vec h_1 s s_1 t t_1 \rangle n_1,
\label{se1}
\ee
where $\bar V_{2N}$ is the anti-symmetrized potential matrix element, 
$n_{p,n} = \theta(k_f(1 \mp \delta_{np})^{1/3}-|\vec h_1|)$ is the 
occupation probability, and the sum is taken over the 
momentum $\vec h_1$, spin $s_1$, and isospin $t_1$ of the intermediate hole state. 
In the present work we compute the exact dependence of $\Sigma^{(1),2N}_{p,n}(q;k_f,\delta_{np})$
on $\delta_{np}$ and extract the linear and quadratic terms numerically.

\begin{center}
\begin{figure}
\begin{center}
\includegraphics[height=5.9cm]{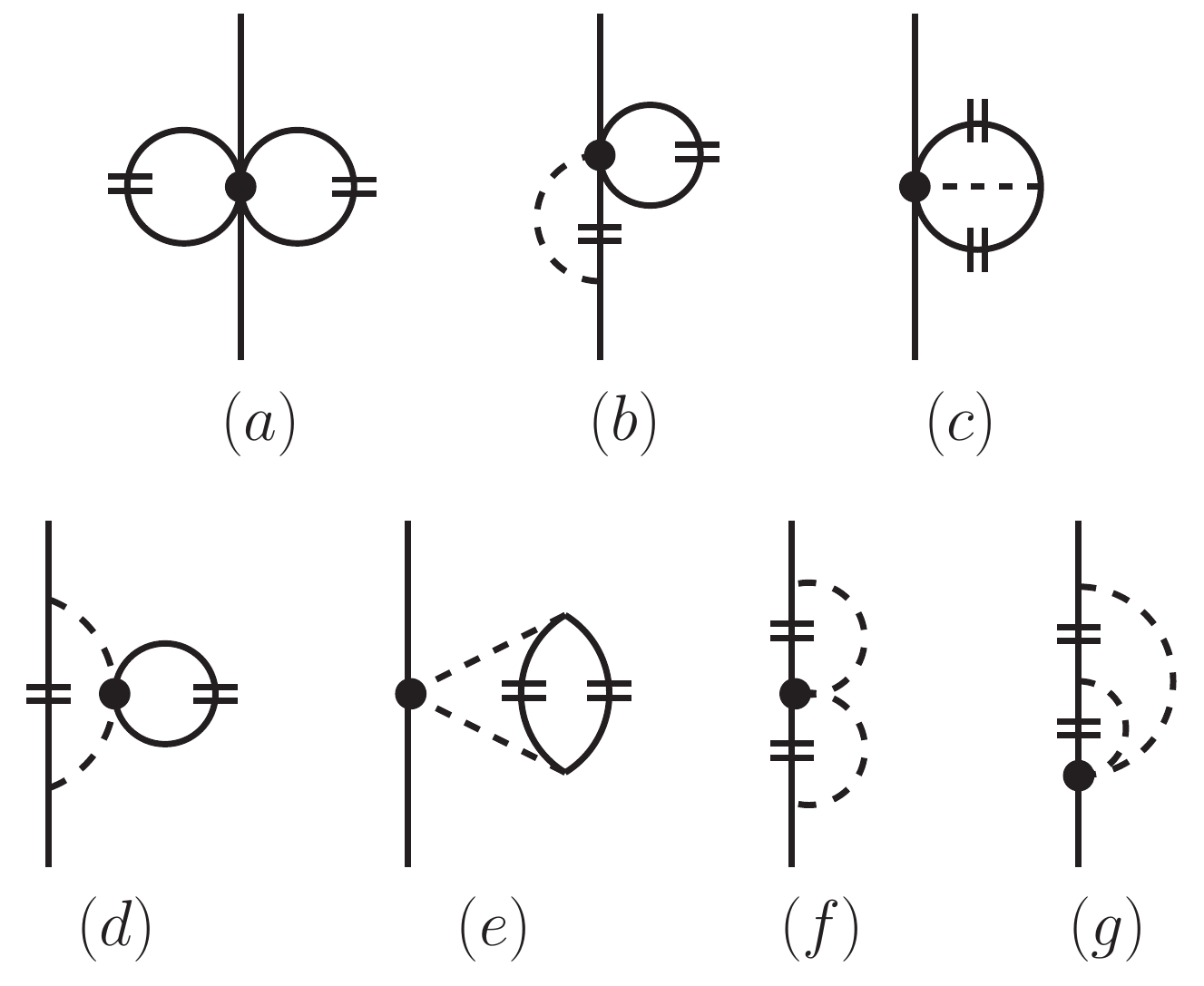}
\end{center}
\vspace{-.2cm}
\caption{Contributions to the Hartree-Fock single-particle potential from the N$^2$LO 
chiral three-nucleon force. The large dots represent vertices 
proportional to the low-energy constants $c_1,c_3,c_4,c_D,c_E$, and the short 
double-line indicates a medium insertion: $-2\pi \delta(k_0) \theta (k_f^{p,n} - |\vec k|)$.}
\label{op3nf}
\end{figure}
\end{center}

The Hartree-Fock contributions from three-body forces are obtained by summing two particles over the
filled states in the fermi sea:
\begin{eqnarray}
&&\hspace{-.15in}\Sigma^{(1),3N}_t(q;k_f,\delta_{np}) \\
&&\hspace{-.15in}= \frac{1}{2}\sum_{12} \langle \vec q \, \vec h_1\vec h_2; s s_1s_2; t t_1t_2 |
 \bar V_{3N} | \vec q \, \vec h_1\vec h_2; s s_1s_2; t t_1t_2 \rangle n_1 n_2.
 \nonumber
\label{se31}
\end{eqnarray}
The next-to-next-to-leading order (N$^2$LO) chiral three-body force consists of three terms,
whose diagrammatic contributions to the nucleon self-energy are shown in Fig.\ \ref{op3nf}.
The chiral three-nucleon contact force is proportional to the low-energy constant $c_E$:
\begin{equation}
V_{3N}^{(\rm ct)} = \sum_{i\neq j\neq k} \frac{c_E}{2f_\pi^4 \Lambda_\chi}
{\vec \tau}_i \cdot {\vec \tau}_j\, ,
\label{3n3}
\end{equation}
where $\Lambda_{\chi} = 700$\,MeV and $f_\pi = 92.4$ MeV.
The one-pion exchange three-body force proportional to the low-energy 
constant $c_D$ has the form
\begin{equation}
V_{3N}^{(1\pi)} = -\sum_{i\neq j\neq k} \frac{g_A c_D}{8f_\pi^4 \Lambda_\chi} 
\frac{\vec{\sigma}_j \cdot \vec{q}_j}{\vec{q_j}^2+m_\pi^2} \vec{\sigma}_i \cdot
\vec{q}_j \, {\vec \tau}_i \cdot {\vec \tau}_j \, ,
\label{3n2}
\end{equation}
where $g_A=1.29$ and $m_{\pi} = 138$ MeV. 
Finally, the two-pion exchange component with vertices proportional to $c_{1,3,4}$ is given by
\be
V_{3N}^{(2\pi)} = \sum_{i\neq j\neq k} \frac{g_A^2}{8f_\pi^4} 
\frac{\vec{\sigma}_i \cdot \vec{q}_i \, \vec{\sigma}_j \cdot
\vec{q}_j}{(\vec{q_i}^2 + m_\pi^2)(\vec{q_j}^2+m_\pi^2)}
F_{ijk}^{\alpha \beta}\tau_i^\alpha \tau_j^\beta,
\label{3n1}
\ee
where the isospin tensor is  
\begin{equation}
F_{ijk}^{\alpha \beta} = \delta^{\alpha \beta}\left (-4c_1m_\pi^2
 + 2c_3 \vec{q}_i \cdot \vec{q}_j \right ) + 
c_4 \epsilon^{\alpha \beta \gamma} \tau_k^\gamma \vec{\sigma}_k
\cdot \left ( \vec{q}_i \times \vec{q}_j \right ).
\label{3n4}
\end{equation}
The low-energy constants $c_{1,3,4}$ have been fitted (within the empirical uncertainties imposed 
by $\pi N$ scattering) to nucleon-nucleon scattering 
phase shifts \cite{machleidt11}, while the $c_D$ and $c_E$ constants have been fitted to reproduce
the binding energies of $^3$H and $^3$He as well as the $\beta$-decay lifetime of
$^3$H \cite{coraggio14}.

At the Hartree-Fock level, the first-order terms in $\delta_{np}$ arising from contact 
three-nucleon forces proportional to the low-energy constants $c_E$ and $c_D$ are given by
\begin{eqnarray}
U_I(q,k_f) &=& {c_E k_f^6\over 6\pi^4 f_\pi^4 \Lambda_\chi} + {g_Ac_Dm_\pi^6u^3\over 3(2\pi f_\pi)^4\Lambda_\chi}
\left \{2u-2u^3 \right .\\ \nonumber
&& \hspace{-.75in} -\arctan 2u - \arctan(u+x)-\arctan(u-x)  \\ \nonumber
&& \hspace{-.75in} + {1\over 4u}
\ln(1+4u^2) +{3+5u^2-3x^2 \over 12x} \ln{1+(u+x)^2 \over 1+(u-x)^2}\bigg\}\,,
\end{eqnarray}
where $k_f^3 = 3\pi^2(\rho_n+\rho_p)/2$, $x=q/m_\pi$ and $u = k_f/m_\pi$.
The $2\pi$-exchange Hartree diagrams proportional to $c_{1,3}$ give rise to the isovector
optical potential strength function
\begin{eqnarray} \nonumber
U_I(q,k_f) &=& {g_A^2 m_\pi^6 u^5 \over 18\pi^4f_\pi^4}\bigg\{ c_3 u 
+ {c_1-c_3 \over 2x}\ln{1+(u+x)^2\over 1+(u-x)^2}
 \\
&+& {(c_3-2c_1)u \over [1+(u+x)^2][1+(u-x)^2]}  \bigg\}
\,,
\end{eqnarray}
while the $2\pi$-exchange Fock diagrams proportional to $c_{1,3,4}$ yield
\begin{eqnarray} \nonumber
U_I(q,k_f) &=&{g_A^2 m_\pi^6 u \over 9(4\pi f_\pi)^4 x^2} \bigg\{
-6c_1\Big[H(x,u)\, \partial_u H(x,u) \\ \nonumber
&&\hspace{-.75in}+H(u,u)\, \partial_x H(u,x)\Big]
+(2c_4-c_3)\Big[G(x,u)\, \partial_u G(x,u) \\ \nonumber
&&\hspace{-.75in} +G(u,u)\,
\partial_xG(u,x)\Big] -2(c_3+c_4)\Big[I(x,u)\, \partial_u I(x,u) \\ \nonumber
&&\hspace{-.75in}+I(u,u)\, \partial_x I(u,x)\Big] +\int_0^u\!d\xi
\Big[ 18c_1 \partial_u H(\xi,u) \, \partial_x H(\xi,x) \\ \nonumber
&&\hspace{-.75in}+(3c_3+2c_4) \partial_u
G(\xi,u)\, \partial_x G(\xi,x)+2(3c_3-c_4) \\
&& \hspace{-.75in} \times \partial_u
I(\xi,u) \,\partial_x I(\xi,x) \Big] \bigg\}\,, 
\end{eqnarray}
with auxiliary functions
\begin{eqnarray}\nonumber
\hspace{-.2in}G(x,u) &=& {4u x\over 3}(2u^2-3)+4x \big[\arctan(u+x)\\ 
&&\hspace{-.75in}+\arctan(u-x)\big] +(x^2-u^2-1)\ln{1+(u+x)^2 \over 1+(u-x)^2} \,, 
\label{aux1}
\end{eqnarray}
\begin{eqnarray}\nonumber
\hspace{-.3in}H(x,u) &=&  u(1+u^2+x^2)-{1\over 4x}\big[1+(u+x)^2\big] \\ 
&\times& \big[1+(u-x)^2\big] \ln{1+(u+x)^2 \over 1+(u-x)^2}\,,
\label{aux2}
\end{eqnarray}
\begin{eqnarray}
I(x,u) &=& {u x\over 6}(8u^2+3x^2)-{u\over 2x}(1+u^2)^2+
{1\over 8}\bigg[{(1+u^2)^3\over x^2} \nonumber \\ 
&&\hspace{-.55in} -x^4+(1-3u^2)(1+u^2-x^2)
\bigg] \ln{1+(u+x)^2\over 1+(u-x)^2}\,,
\label{aux3}
\end{eqnarray}
The terms second-order in $\delta_{np}$ are relatively small, and 
the explicit expressions are given in the Appendix.

Finally, we consider the second-order
perturbative contribution from two and three-body forces, $U_{2N+3N}^{(2)}$,
which is approximated by employing a density-dependent NN interaction constructed 
from $V_{3N}$ as described in Refs.\ \cite{holt09,holt10,hebeler10}.
The second-order contributions, shown in Figs.\ \ref{op2nf}(a) and \ref{op2nf}(b), 
are given by
\begin{eqnarray}
&&\hspace{-.3in}\Sigma^{(2a),2N}_t(q,\omega;k_f,\delta_{np})  \nonumber \\
&&\hspace{-.15in}= \frac{1}{2}\sum_{123} \frac{| \langle \vec p_1 \vec p_3 s_1 s_3 t_1 
t_3 | \bar V_{2N}^{\rm eff} | \vec q \, \vec h_2 s s_2 t t_2 \rangle |^2}{\omega + \epsilon_2 - \epsilon_1
-\epsilon_3 + i \eta} \bar n_1 n_2 \bar n_3 \nonumber \\
&&\times (2\pi)^3 \delta(\vec p_1 + \vec p_3 - \vec q - \vec h_2),
\label{op2ac}
\end{eqnarray}

\begin{eqnarray}
&&\hspace{-.3in}\Sigma^{(2b),2N}_t(q,\omega;k_f,\delta_{np})  \nonumber \\
&&\hspace{-.15in}= \frac{1}{2}\sum_{123} \frac{| \langle \vec h_1 \vec h_3 s_1 s_3 t_1 
t_3 | \bar V_{2N}^{\rm eff} | \vec q \, \vec p_2 s s_2 t t_2 \rangle |^2}{\omega + \epsilon_2 - \epsilon_1
- \epsilon_3 - i \eta} n_1 \bar n_2 n_3 \nonumber \\
&&\times (2\pi)^3 \delta(\vec h_1 + \vec h_3 - \vec q - \vec p_2).
\label{op2bd}
\end{eqnarray}
where $\bar n_k = 1-n_k$ and $\bar V_{2N}^{\rm eff}$ is the sum of the free-space nucleon-nucleon
potential $\bar V_{2N}$ and the density-dependent two-body force $\bar V_{2N}^{\rm med}$ 
\cite{holt09,holt10}. All single-particle energies are computed self-consistently:
\begin{eqnarray}
e_{p,n}(q) &=& q^2/2M + \Sigma^{(1),2N}_{p,n}(q) \\ \nonumber
&+& \Sigma^{(1),3N}_{p,n}(q) + {\rm Re}\, \Sigma^{(2)}_{p,n}(q,e(q)).
\end{eqnarray}
The expressions are computed for arbitrary isospin asymmetry, and the linear and quadratic
terms in $\delta_{np}$ are extracted numerically.

\begin{figure}
\includegraphics[width=8.7cm]{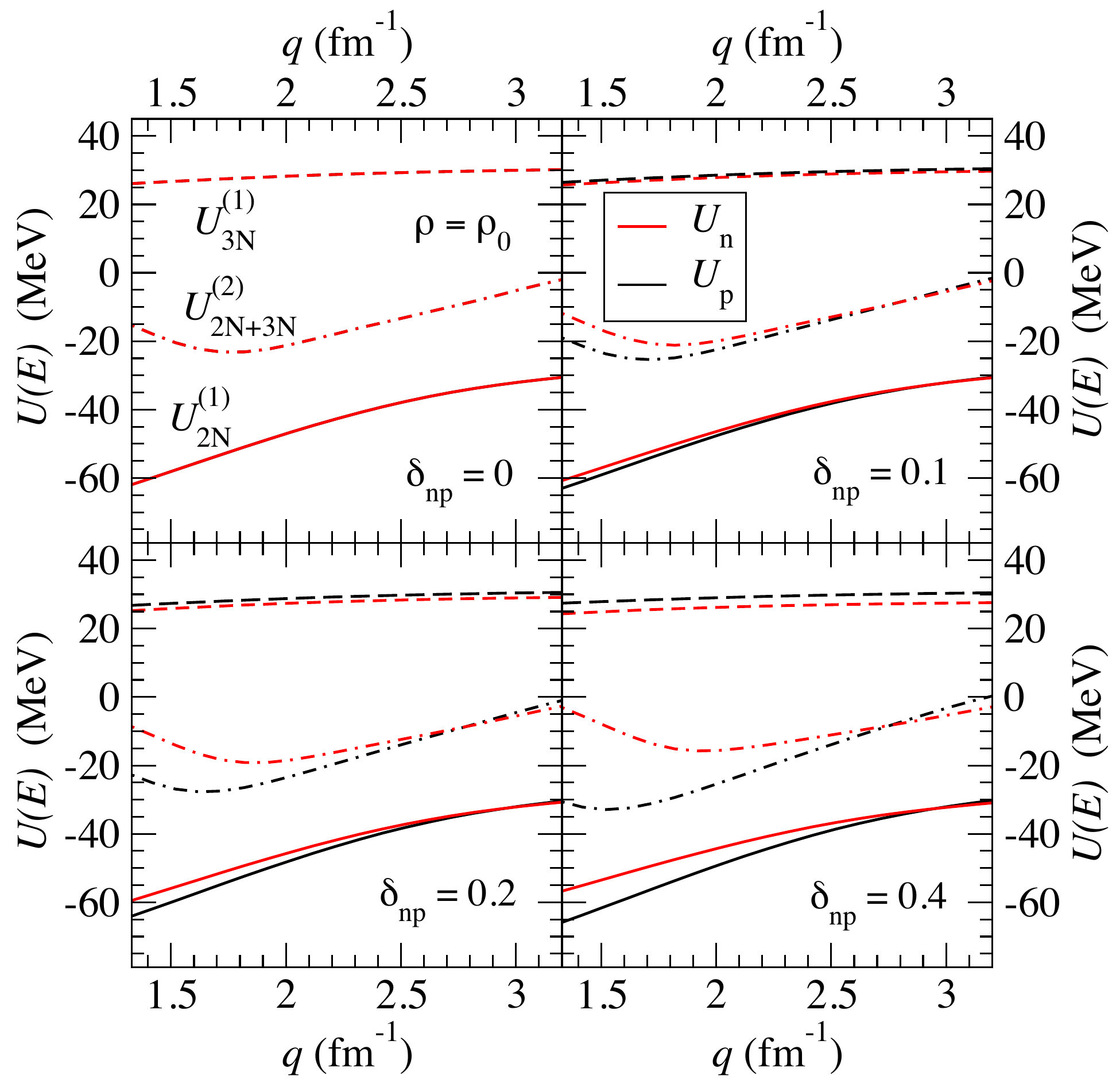}
\caption{(color online) Contributions to the real part of the proton and neutron optical model potentials 
from the n3lo450 chiral two- and three-body forces as a function of the momentum $q$ and isospin asymmetry 
$\delta_{np}$ at nuclear matter saturation density $\rho_0$. }
\label{OP2N3N}
\end{figure}

We show in Fig.\ \ref{OP2N3N} the various contributions to the real part of the single-particle
potential for protons and neutrons in asymmetric matter at a density of $\rho_0$ computed
from the n3lo450 chiral NN potential. We notice that when employing the n3lo450 low-momentum 
chiral nuclear potential (and also the n3lo414 potential), the second-order 
perturbative contribution is 
always less than the Hartree-Fock contribution, in contrast to the behavior observed in
Ref.\ \cite{holt13b} using the n3lo500 potential. We find that three-nucleon forces in general
enhance isospin inversion due to the repulsive character of three-body forces in homogeneous matter.

\begin{figure}
\includegraphics[width=8.5cm]{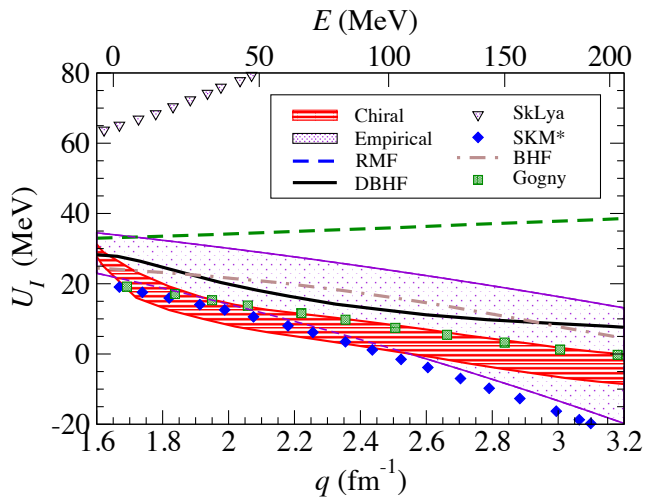}
\caption{(color online) Energy dependence of the isovector real optical model potential at saturation density
from chiral effective field theory. Shown for comparison are the predictions of other microscopic,
semi-microscopic, and phenomenological models (see Ref.\ \cite{dalen05}).}
\label{ReIv}
\end{figure}

The magnitude and energy dependence of the real isovector part of the 
optical potential are poorly constrained by experiment.
From Refs.\ \cite{becchetti69,patterson76,rapaport79,rapaport79b,varner91,koning03,li04} 
one finds that the magnitude is expected to decrease with energy according to
$U_I = ( 28 \pm 6 )\,{\rm MeV} - ( 0.15 \pm 0.05 ) E,$
which we show as the empirical band in Fig.\ \ref{ReIv}. The chiral effective field theory
prediction shown in Fig.\ \ref{ReIv} is consistent with the empirical constraints but 
has significantly smaller uncertainties, 
typically of about 5-10 MeV over a wide range of scattering energies. The region for isospin inversion 
is predicted to lie in the range $E_{\rm inv} = 155 \pm 45$\,MeV, and the two low-momentum interactions
alone would give a much narrower region of $E_{\rm inv} = 120 \pm 10$\,MeV. The results from other
microscopic many-body calculations are shown, including Brueckner-Hartree-Fock 
``BHF'' \cite{zuo05} and Dirac-Brueckner-Hartree-Fock ``DBHF'' \cite{dalen05},
as well as semi-microscopic mean field models: ``RMF'' \cite{gaitanos04}, ``Gogny'' \cite{kleban02}, 
``SKM*'' \cite{bartel82}, and ``SkLya'' \cite{cochet04}. We refer the reader to Ref.\ \cite{dalen05} for 
additional details and analysis.

In addition to variations in the resolution scale, also the order-by-order convergence 
\cite{sammarruca15,epelbaum15} in the chiral expansion provides an estimate of the 
theoretical uncertainty and associated cutoff artifacts. For this purpose we
have computed as well the optical potential from the NLO (next-to-leading order) and 
N2LO chiral potentials with
cutoffs $\Lambda = 450$ and 500 MeV \cite{sammarruca15}. At threshold the uncertainties
for both sets of cutoffs are similar: $U_I^{450}(0) = 32 \pm 6$ MeV and 
$U_I^{500}(0) = 31 \pm 6$ MeV.  At the isospin inversion energy the uncertainties
are slightly reduced: $U_I^{450}(E_{inv}) = 0 \pm 5$ MeV and $U_I^{500}(E_{inv}) = 0 
\pm 4$ MeV. Accounting for these uncertainties would not qualitatively alter the error bands
shown in Fig.\ \ref{ReIv}, except at low scattering energies. We also note that beyond an energy of 
$E \simeq 200$\,MeV, significant artifacts were observed in the calculation of the optical
potential from the n3lo414 potential, suggesting a breakdown in the chiral effective field
theory expansion.

\begin{figure}[t]
\includegraphics[width=8.5cm]{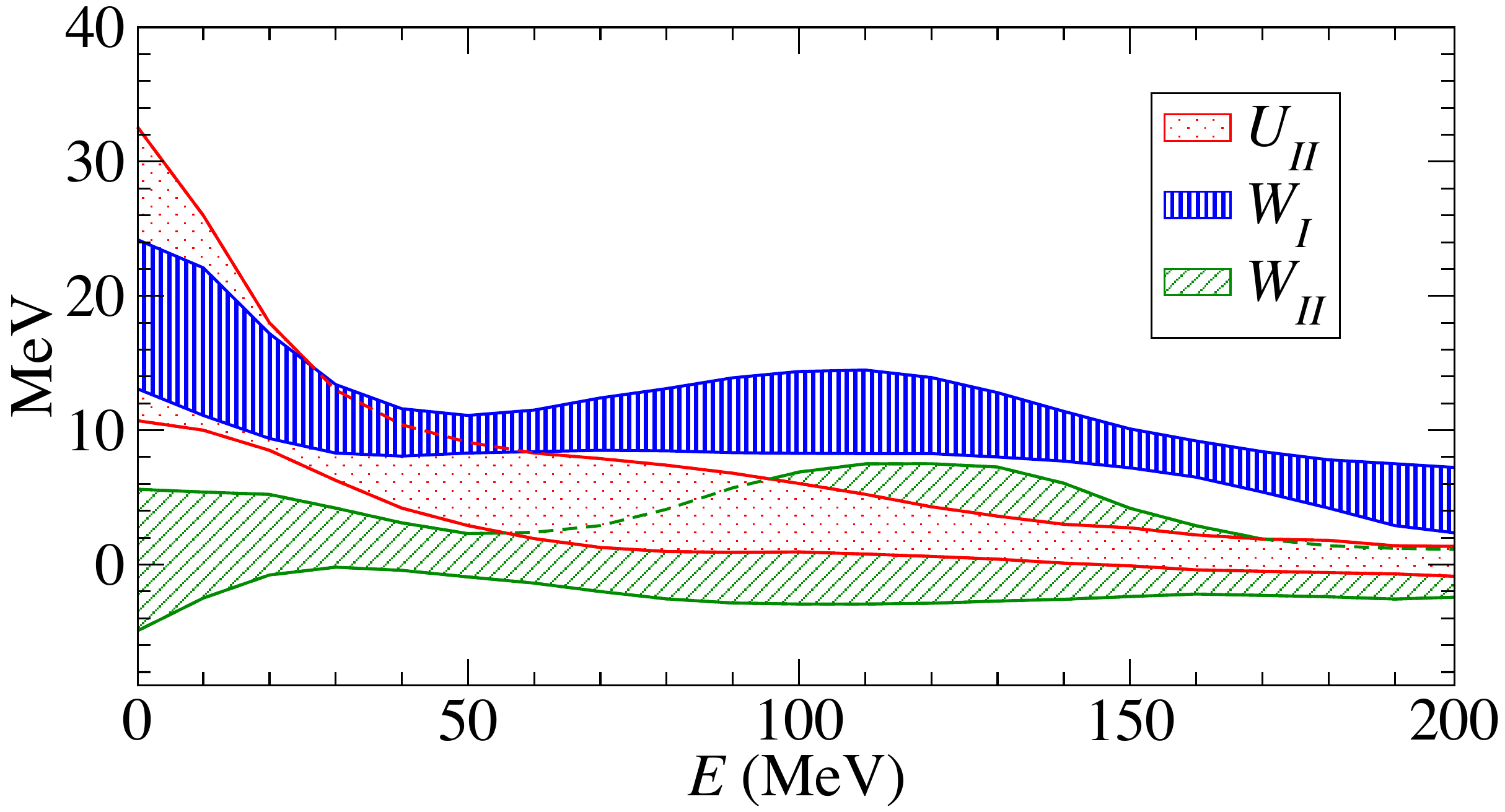}
\caption{Energy dependence of the isovector imaginary optical potential at saturation density 
from chiral two- and three-body forces. Also shown are the subleading $\delta_{np}^2$ contributions 
to both the real and imaginary potentials.}
\label{UII}
\end{figure}

In Fig.\ \ref{UII} we show the subleading contribution $U_{II}$ to the real optical potential
as a function of projectile energy. For low-energy scattering on neutron-rich targets we can
expect an isoscalar shift of roughly ($15-20$\,MeV)$\delta_{np}^2$, while
for energies greater than $E>100$ MeV the quadratic term is consistent with zero. 
The latter observation has the important consequence that extrapolations of phenomenological
optical potentials into the neutron-rich region of the nuclear chart can be valid for energies beyond
$E>100$\,MeV. This is supported by a similar feature in the volume imaginary component, where
in Fig.\ \ref{UII} we see that the quadratic $\delta_{np}^2$ term $W_{II}$ is consistent with zero for all
scattering energies considered. The Lane parametrization of the volume imaginary optical 
potential strength therefore provides an excellent approximation to the true isospin asymmetry
dependence over a large range of energies. 
We note that since no isovector component for the volume imaginary contribution 
could be extracted from the most recent analyses in Refs.\ \cite{varner91,koning03} due to the 
uncertainties in the scattering data at large energies, our result is a
prediction that may be verified at RIB facilities. 

\begin{figure}[t]
\includegraphics[width=8.5cm]{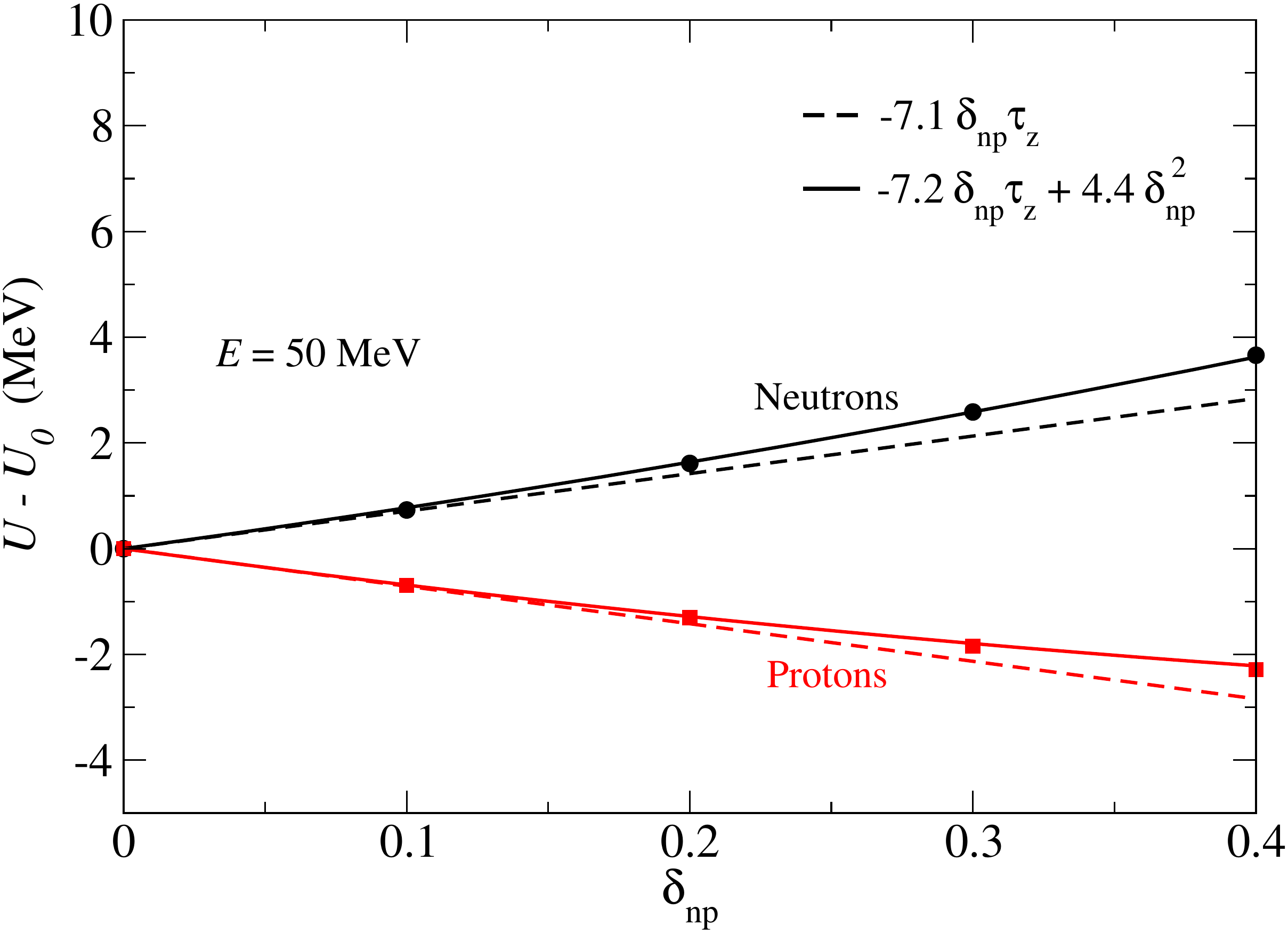}
\caption{Isospin asymmetry dependence of the proton and neutron real optical potentials in infinite 
nuclear matter at saturation density $\rho_0$. The squares and circles are the calculated results, while
the dashed and solid lines are linear and quadratic fits, respectively. A scattering energy of $E=50$\,MeV 
has been chosen.}
\label{qual}
\end{figure}

Finally, we consider the quality of fitting the isospin asymmetry dependence of the proton
and neutron real optical potentials up to quadratic $\delta_{np}^2$ terms. In Fig.\ \ref{qual} we 
show as a representative example the isospin asymmetry dependence 
of the optical potential for a scattering energy of 50\,MeV employing the n3lo450 chiral 
nuclear potential. For values of the isospin asymmetry up to $\delta_{np} = 0.1$ the leading 
contribution is isovector in character and the Lane parametrization works well.
Even at $\delta_{np} = 0.2$ the second-order $\delta_{np}^2$ isoscalar contribution becomes 
evident. At the highest value of $\delta_{np} = 0.4$, the isoscalar $\delta_{np}^2$ term gives a 
contribution to the optical potential that is about 1/4 that of the isovector contribution.

The present work lays the foundation for improved modeling of nucleon-nucleus scattering 
away from the valley of stability. A strong isovector component to the imaginary part of the optical
potential, as found in the present study, can inhibit radiative neutron-capture cross sections on exotic 
nuclei \cite{goriely07} and strongly influence $r$-process nucleosynthesis in cooler environments 
such as the tidally ejected matter in neutron star mergers. The isovector real part of the single-particle 
potential (together with the isoscalar $\delta_{np}^2$ terms important at low energies) may be more 
relevant for neutron star inner crusts, where a lattice (or pasta structures) of neutron-rich 
nuclei interact with a background of free neutrons. At the higher energies attained at next-generation
radioactive beam facilities, our predicted sign change in the isovector real optical potential in the 
energy range $110<E_{inv}<200$\,MeV as well as the strength of the isovector imaginary 
optical potential $W_I \simeq 8-12$\, MeV can be tested.

Work supported in part by US DOE Grant No.\ DE-FG02-97ER-41014, the BMBF, the DFG cluster of 
excellence Origin and Structure of the Universe, by the DFG, NSFC (CRC110).

\section{Appendix: $\bm{\delta_{np}^2}$ contributions from three-nucleon forces}

Here we present results for the quadratic $\delta_{np}^2$ corrections to the nucleon-nucleus
optical potential from three-body forces at the 
Hartree-Fock level, which are found to be 
significantly smaller than the leading $\delta_{np}$ terms.
The short-distance contact term contribution, shown in Fig.\ \ref{op3nf}(a), has the form
\begin{equation}
U_{I\!I}(q,k_f) = {c_E k_f^6 \over 12\pi^4 f_\pi^4 \Lambda_\chi}.
\end{equation}
The two diagrams from the one-pion exchange three-body force, Figs.\ \ref{op3nf}(b) and 
\ref{op3nf}(c), yield
\begin{eqnarray} \nonumber
U_{I\!I}(q,k_f) &=& {g_A c_D m_\pi^6 u^5 \over 18(2\pi f_\pi)^4
\Lambda_\chi}\bigg\{{8u^2+3\over 4u^3}\ln(1+4u^2) -6u \\ \nonumber
&-& {3\over u} +{u+x+x^{-1} \over 1+(u+x)^2} +{u-x-x^{-1} \over 1+(u-x)^2} \\
&+&{3\over 2x}\ln{1+(u+x)^2 \over 1+(u-x)^2}\bigg\}\,.
\end{eqnarray}
The Hartree diagrams, Figs.\ \ref{op3nf}(d) and 
\ref{op3nf}(e), from the two-pion exchange three-body force proportional
to $c_1$ and $c_3$ give
\begin{eqnarray} \nonumber 
U_{I\!I}(q,k_f) &=& {g_A^2 m_\pi^6 u^5 \over 9(2\pi f_\pi)^4}
\bigg\{4c_3 u+{6 \over u}(3c_3-4c_1) \\ \nonumber
&&\hspace{-.6in}+{8u(2c_1-c_3)\over 1+4u^2} + \bigg[{8 \over u}(c_1-c_3) 
+{3 \over 2u^3}(4c_1-3c_3)\bigg] \\ \nonumber
&&\hspace{-.6in}\times \ln(1+4u^2)+{2\over x}(c_1-c_3)\ln{1+(u+x)^2 \over 1+(u-x)^2} \\ \nonumber
&&\hspace{-.6in}+{8u(c_3-c_1)(1+x^2-u^2) \over [1+(u+x)^2][1+(u-x)^2]} \\
&&\hspace{-.6in}+{16u^3(c_3 -2c_1)
(1+u^2-x^2)\over [1+(u+x)^2]^2[1+(u-x)^2]^2} \bigg\}\,.
\end{eqnarray}
The Fock diagrams, Figs.\ \ref{op3nf}(f) and 
\ref{op3nf}(g), are split into two parts depending on $c_1$ and $c_{3,4}$:
\begin{eqnarray} \nonumber
U_{I\!I}(q,k_f) &=&{g_A^2c_1m_\pi^6u\over 3(4\pi f_\pi)^4x^2}\bigg\{
H(x,u)\Big[u\, \partial^2_uH(x,u) \\ \nonumber
&&\hspace{-.75in} -2\partial_u H(x,u)\Big] +u \big[\partial_u
H(x,u)\big]^2 + \partial_xH(u,x)\Big[ {u\over 3}(3\partial_x \\ \nonumber
&&\hspace{-.75in}-2 \partial_u)H(x,u)\big|_{x=u}-2 H(u,u)\Big]+u H(u,u)\,\partial_u\partial_x H(u,x) \\
&& \hspace{-.75in} + \int_0^u\!d\xi\, \partial_x H(\xi,x) \Big[ u\,
\partial^2_u H(\xi,u) -2 \partial_u H(\xi,u) \Big] \bigg\}\, ,
\end{eqnarray}

\begin{eqnarray} \nonumber
U_{I\!I}(q,k_f) &= &{g_A^2 m_\pi^6u\over 18
(4\pi f_\pi)^4x^2}\bigg\{{u \over 3}(3c_3+2c_4) \big[\partial_u G(x,u)\big]^2 \\ \nonumber
&&\hspace{-.8in}+(2c_4-c_3)G(x,u) \Big[2 \partial_u G(x,u)- u\,\partial^2_u
G(x,u)\Big] +{2u \over 3}(3c_3 \\ \nonumber
&& \hspace{-.8in} -c_4) \big[\partial_u I(x,u)\big]^2+2(c_3+c_4)
I(x,u)\Big[u\, \partial^2_u I(x,u) \\ \nonumber
&&\hspace{-.8in}-2\partial_u I(x,u)\Big]
+(c_3-2c_4)\bigg[ \partial_x G(u,x)\Big({u\over 3}(3\partial_x G(x,u) \big |_{x=u} \\ \nonumber
&& \hspace{-.8in} -2\partial_u G(x,u) \big |_{x=u}-2G(u,u)\Big) +u G(u,u) \,\partial_u
\partial_x G(u,x) \\ \nonumber
\end{eqnarray}
\begin{eqnarray} \nonumber
&& + \int_0^u\!d\xi\,\partial_x G(\xi,x) \Big(u \,
\partial^2_u G(\xi,u)-2 \partial_u G(\xi,u) \Big)\bigg]
+2(c_3 \\ \nonumber
&&  +c_4)\bigg[ \partial_x I(u,x)\Big({u\over 3}
(3\partial_x-2\partial_u) I(x,u)\big|_{x=u}-2I(u,u)\Big) \\ \nonumber
&&+u I(u,u) \,\partial_u\partial_x I(u,x) + \int_0^u\!d\xi\,\partial_x I(\xi,x)
\Big(u \, \partial^2_u I(\xi,u) \\ 
&& -2 \partial_u I(\xi,u) \Big)\bigg] \bigg\}\,,
\end{eqnarray}
with auxiliary functions defined in Eqs.\ (\ref{aux1})-(\ref{aux3}).


\bibliographystyle{apsrev4-1}
%


\end{document}